\documentclass{article}
\usepackage{makeidx}
\usepackage{graphicx}
\usepackage{amsmath}

\input{tcilatex}

\begin{document}

\title{Quantum mechanics gives stability to a Nash equilibrium.}
\author{A. Iqbal and A.H. Toor \\
Electronics Department, Quaid-i-Azam University, \\
Islamabad, Pakistan.\\
email: qubit@isb.paknet.com.pk}
\maketitle

\begin{abstract}
We consider a slightly modified version of the Rock-Scissors-Paper (RSP)
game from the point of view of evolutionary stability. In its classical
version the game has a mixed Nash equilibrium (NE) not stable against
mutants. We find a quantized version of the RSP game for which the classical
mixed NE becomes stable.
\end{abstract}

\section{Introduction}

Long played as a children's pastime, or as an odd-man-out selection process,
the Rock-Scissors-Paper (RSP) is a game for two players typically played
using the players' hands. The two players opposite each others, tap their
fist in their open palm three times (saying Rock, Scissors, Paper) and then
show one of three possible gestures. The Rock wins against the scissors
(crushes it) but looses against the paper (is wrapped into it). The Scissors
wins against the paper (cuts it) but looses against the rock (is crushed by
it). The Paper wins against the rock (wraps it) but looses against the
scissors (is cut by it).

In a slightly modified version of the RSP game both players get a small
premium $\epsilon $\ for a draw. This game can be represented by the
following payoff matrix

\begin{equation}
\left( 
\begin{array}{cccc}
& R & S & P \\ 
R & -\epsilon & 1 & -1 \\ 
S & -1 & -\epsilon & 1 \\ 
P & 1 & -1 & -\epsilon
\end{array}
\right)  \label{matrix}
\end{equation}
where $-1<\epsilon <0$. The matrix of the usual game is obtained when $%
\epsilon $ is zero in the matrix (\ref{matrix}).

One cannot win if one's opponent knew which strategy was going to be picked.
For example, picking Rock consistently all the opponent needs to do is pick
Paper and s/he would win. Players find soon that in case predicting
opponent's strategy is not possible the best strategy is to pick Rock,
Scissors, or Paper at random. In other words, the player selects Rock,
Scissors, or Paper with a probability of $\frac{1}{3}$. In case opponent's
strategy is predictable picking a strategy at random with a probability of $%
\frac{1}{3}$ is not the best thing to do unless the opponent is doing the
same \cite{prestwich}.

We explore Evolutionarily Stable Strategies (ESSs) in quantized RSP game.
Originally defined by Smith and Price \cite{smith}\ as a behavioral
phenotype an ESS cannot be invaded by a mutant strategy when a population is
playing it. A mutant strategy does things in different ways than most of a
population does. Smith and Price considered a symmetric game where the
players are anonymous. Let $P[u,v]$ be the payoff to a player playing $u$
against the player playing $v$. Strategy $u$ is an ESS if for any
alternative strategy $v$, the following two requirements are satisfied

\begin{equation}
P[u,u]\geq P[v,u]  \label{ess1}
\end{equation}
and in the case $P[u,u]=P[v,u]$

\begin{equation}
P[u,v]>P[v,v]  \label{ess2}
\end{equation}
Requirement (\ref{ess1}) is in fact the Nash requirement and says that no
single individual can gain by unilaterally changing her strategy from $u$ to 
$v$. An ESS is in fact a stable NE in a symmetric game and its stability is
against a small group of mutants \cite{smith, weibull}.

A straight analysis of the modified RSP game of matrix (\ref{matrix}) shows
that playing each of the three different pure strategies with a fixed
equilibrial probability $\frac{1}{3}$ constitutes a mixed NE. However it is
not an ESS because $\epsilon $ is negative \cite{weibull}.

In an earlier paper \cite{iqbal} we showed that in quantized version of
certain asymmetric games between two players it is possible to make appear
or disappear an ESS that is a pure strategy pair NE by controlling the
initial state used to play the game. Because a classical game is embedded in
its quantized form, therefore, it is possible that a pure strategy pair NE
remains intact in both classical and certain quantized form of the same game
but is an ESS in only one form. Later we presented an example \cite{iqbal1}
of a symmetric game between two players for which a pure strategy NE is an
ESS in classical version of the game but not so in a quantized form even
when it remains a NE in both versions. This is more relevant because the
idea of an ESS was originally defined for symmetric contests. We also showed 
\cite{iqbal2} that mixed strategy ESSs can be related to entanglement and
can be affected by quantization for three player games. However it is not
the case for two player games when the quantum state is in a simpler form
proposed by Marinatto and Weber (M\&W) \cite{marinatto} in their scheme to
quantize a two player game in normal form.

M\&W \cite{marinatto} expanded on the scheme proposed by Eisert et al \cite
{eisert} for the game of Prisoner's Dilemma. Eisert et al showed that
dilemma doesn't exist in a quantum version of the game. Their motivation of
M\&W was to remove the need of an unentangling gate in Eisert's scheme \cite
{marinattoP, Benjamin}. In our effort to extend the ideas of evolutionary
game theory toward quantum games we found M\&W's scheme more suitable for
following reasons:

(a). In usual setup of a symmetric bi-matrix evolutionary game two pure
strategies are assumed such that players can play a mixed strategy by their
probabilistic combination. In a similar way players in M\&W's scheme can
play a mixed strategy by applying two unitary operators in their possession
with classical probabilities.

(b): The usual definition of `fitness' of a mixed strategy in evolutionary
games \cite{prestwich}\ can be given a straightforward extension in M\&W's
scheme \cite{iqbal}. It is done when in the quantum game players use only
one unitary operator out of the two.

(c): The theory of ESSs in evolutionary game theory is developed mostly for
situations when players are anonymous and possess a discrete number of pure
strategies. We find that ESS idea can be extended towards quantum settings
more easily in M\&W's scheme than in Eisert's scheme involving a continuum
of pure strategies players have option to play. The idea of an ESS as a
stable equilibrium is confronted with problems when players have an access
to a continuum of pure strategies \cite{oechssler}.

In this paper we want to extend our previous results regarding effects of
quantization on evolutionary stability for a modified version of the RSP
game. This game is different because now classically each player possesses
three pure strategies instead of two. A classical mixed NE exists that is
not an ESS. Our motivation is to explore the possibility that the classical
mixed NE becomes an ESS for some initial entangled quantum state. We show
that such a quantum state not only exists but also easy to find.

\section{Quantized RSP game}

Using simpler notation: $R\sim 1,$ $S\sim 2,$ $P\sim 3$ we quantize this
game via M\&W's scheme \cite{marinatto}. We allow the two players are in
possession of three unitary operators $I,C$ and $D$ defined as

$I\left| 1\right\rangle =\left| 1\right\rangle \qquad C\left| 1\right\rangle
=\left| 3\right\rangle \qquad D\left| 1\right\rangle =\left| 2\right\rangle $

$I\left| 2\right\rangle =\left| 2\right\rangle \qquad C\left| 2\right\rangle
=\left| 2\right\rangle \qquad D\left| 2\right\rangle =\left| 1\right\rangle $

$I\left| 3\right\rangle =\left| 3\right\rangle \qquad C\left| 3\right\rangle
=\left| 1\right\rangle \qquad D\left| 3\right\rangle =\left| 3\right\rangle $

where $C^{\dagger }=C=C^{-1}$ and $D^{\dagger }=D=D^{-1}$ and $I$ is
identity operator. We also start from a general payoff matrix for two
players each having three strategies

\begin{equation}
\left[ 
\begin{array}{cccc}
& 1 & 2 & 3 \\ 
1 & (\alpha _{11},\beta _{11}) & (\alpha _{12},\beta _{12}) & (\alpha
_{13},\beta _{13}) \\ 
2 & (\alpha _{21},\beta _{21}) & (\alpha _{22},\beta _{22}) & (\alpha
_{23},\beta _{23}) \\ 
3 & (\alpha _{31},\beta _{31}) & (\alpha _{32},\beta _{32}) & (\alpha
_{33},\beta _{33})
\end{array}
\right]  \label{genmatrix}
\end{equation}
where $\alpha _{ij}$, $\beta _{ij}$ are payoffs to Alice and Bob
respectively when Alice plays $i$ and Bob plays $j$ and $1\leq i,j\leq 3$.
Suppose Alice and Bob apply operators $C$ $D$ and $I$ with probabilities $p,$
$p_{1}$and $(1-p-p_{1})$ and $q,$ $q_{1}$and $(1-q-q_{1})$ respectively. Let
represent the initial state of the game by $\rho _{in}$. After Alice plays
her strategy the state changes to

\begin{equation}
\overset{A}{\rho _{in}}=(1-p-p_{1})I_{A}\rho _{in}I_{A}^{\dagger
}+pC_{A}\rho _{in}C_{A}^{\dagger }+p_{1}D_{A}\rho _{in}D_{A}^{\dagger }
\end{equation}
The final density matrix after Bob too has played his strategy is

\begin{equation}
\overset{A,B}{\rho _{f}}=(1-q-q_{1})I_{B}\overset{A}{\rho _{in}}%
I_{B}^{\dagger }+qC_{B}\overset{A}{\rho _{in}}C_{B}^{\dagger }+q_{1}D_{B}%
\overset{A}{\rho _{in}}D_{B}^{\dagger }
\end{equation}
This density matrix can be written as

\begin{eqnarray}
\overset{A,B}{\rho _{f}} &=&(1-p-p_{1})(1-q-q_{1})\left\{ I_{A}\otimes
I_{B}\rho _{in}I_{A}^{\dagger }\otimes I_{B}^{\dagger }\right\}
+p(1-q-q_{1})\times  \notag \\
&&\left\{ C_{A}\otimes I_{B}\rho _{in}C_{A}^{\dagger }\otimes I_{B}^{\dagger
}\right\} +p_{1}(1-q-q_{1})\left\{ D_{A}\otimes I_{B}\rho
_{in}D_{A}^{\dagger }\otimes I_{B}^{\dagger }\right\} +  \notag \\
&&(1-p-p_{1})q\left\{ I_{A}\otimes C_{B}\rho _{in}I_{A}^{\dagger }\otimes
C_{B}^{\dagger }\right\} +pq\left\{ C_{A}\otimes C_{B}\rho
_{in}C_{A}^{\dagger }\otimes C_{B}^{\dagger }\right\} +  \notag \\
&&p_{1}q\left\{ D_{A}\otimes C_{B}\rho _{in}D_{A}^{\dagger }\otimes
C_{B}^{\dagger }\right\} +(1-p-p_{1})q_{1}\left\{ I_{A}\otimes D_{B}\rho
_{in}I_{A}^{\dagger }\otimes D_{B}^{\dagger }\right\}  \notag \\
&&+pq_{1}\left\{ C_{A}\otimes D_{B}\rho _{in}C_{A}^{\dagger }\otimes
D_{B}^{\dagger }\right\} +p_{1}q_{1}\left\{ D_{A}\otimes D_{B}\rho
_{in}D_{A}^{\dagger }\otimes D_{B}^{\dagger }\right\}
\end{eqnarray}
The basis vectors of initial quantum state with three pure classical
strategies are $\left| 11\right\rangle ,\left| 12\right\rangle ,\left|
13\right\rangle ,\left| 21\right\rangle ,\left| 22\right\rangle ,\left|
23\right\rangle ,\left| 31\right\rangle ,\left| 32\right\rangle ,$ and $%
\left| 33\right\rangle $. Setting the initial quantum state to following
general form

\begin{eqnarray}
\left| \psi _{in}\right\rangle &=&c_{11}\left| 11\right\rangle +c_{12}\left|
12\right\rangle +c_{13}\left| 13\right\rangle +c_{21}\left| 21\right\rangle
+c_{22}\left| 22\right\rangle +c_{23}\left| 23\right\rangle +  \notag \\
&&c_{31}\left| 31\right\rangle +c_{32}\left| 32\right\rangle +c_{33}\left|
33\right\rangle  \label{IniStat}
\end{eqnarray}
with normalization

\begin{eqnarray}
&&\left| c_{11}\right| ^{2}+\left| c_{12}\right| ^{2}+\left| c_{13}\right|
^{2}+\left| c_{21}\right| ^{2}+\left| c_{22}\right| ^{2}+\left|
c_{23}\right| ^{2}+  \notag \\
&&\left| c_{31}\right| ^{2}+\left| c_{32}\right| ^{2}+\left| c_{33}\right|
^{2}=1
\end{eqnarray}
and writing payoff operators for Alice and Bob as \cite{marinatto}

\begin{eqnarray}
(P_{A,B})_{oper} &=&(\alpha ,\beta )_{11}\left| 11\right\rangle \left\langle
11\right| +(\alpha ,\beta )_{12}\left| 12\right\rangle \left\langle
12\right| +(\alpha ,\beta )_{13}\left| 13\right\rangle \left\langle
13\right| +  \notag \\
&&(\alpha ,\beta )_{21}\left| 21\right\rangle \left\langle 21\right|
+(\alpha ,\beta )_{22}\left| 22\right\rangle \left\langle 22\right| +(\alpha
,\beta )_{23}\left| 23\right\rangle \left\langle 23\right| +  \notag \\
&&(\alpha ,\beta )_{31}\left| 31\right\rangle \left\langle 31\right|
+(\alpha ,\beta )_{32}\left| 32\right\rangle \left\langle 32\right| +(\alpha
,\beta )_{33}\left| 33\right\rangle \left\langle 33\right|
\end{eqnarray}
the payoffs to Alice or Bob can be obtained by taking a trace of $%
(P_{A,B})_{oper}\overset{A,B}{\rho _{f}}$ i.e. \cite{marinatto}

\begin{equation}
P_{A,B}=tr[\left\{ (P_{A,B})_{oper}\right\} \overset{A,B}{\rho _{f}}]
\label{payoff}
\end{equation}
Payoff to Alice, for example, can be written as

\begin{equation}
P_{A}=\Phi \Omega \Upsilon ^{T}  \label{payoffA}
\end{equation}
where $T$ is for transpose and the matrices $\Phi ,$ $\Omega ,$ and $%
\Upsilon $ are

\begin{eqnarray}
\Phi &=&[ 
\begin{array}{ccc}
(1-p-p_{1})(1-q-q_{1}) & p(1-q-q_{1}) & p_{1}(1-q-q_{1})
\end{array}
\notag \\
&& 
\begin{array}{cccccc}
(1-p-p_{1})q & pq & p_{1}q & (1-p-p_{1})q_{1} & pq_{1} & p_{1}q_{1}
\end{array}
]  \notag \\
\Upsilon &=&[ 
\begin{array}{ccccccccc}
\alpha _{11} & \alpha _{12} & \alpha _{13} & \alpha _{21} & \alpha _{22} & 
\alpha _{23} & \alpha _{31} & \alpha _{32} & \alpha _{33}
\end{array}
]  \notag \\
\Omega &=&\left[ 
\begin{array}{ccccccccc}
\left| c_{11}\right| ^{2} & \left| c_{12}\right| ^{2} & \left| c_{13}\right|
^{2} & \left| c_{21}\right| ^{2} & \left| c_{22}\right| ^{2} & \left|
c_{23}\right| ^{2} & \left| c_{31}\right| ^{2} & \left| c_{32}\right| ^{2} & 
\left| c_{33}\right| ^{2} \\ 
\left| c_{31}\right| ^{2} & \left| c_{32}\right| ^{2} & \left| c_{33}\right|
^{2} & \left| c_{21}\right| ^{2} & \left| c_{22}\right| ^{2} & \left|
c_{23}\right| ^{2} & \left| c_{11}\right| ^{2} & \left| c_{12}\right| ^{2} & 
\left| c_{13}\right| ^{2} \\ 
\left| c_{21}\right| ^{2} & \left| c_{22}\right| ^{2} & \left| c_{23}\right|
^{2} & \left| c_{11}\right| ^{2} & \left| c_{12}\right| ^{2} & \left|
c_{13}\right| ^{2} & \left| c_{31}\right| ^{2} & \left| c_{32}\right| ^{2} & 
\left| c_{33}\right| ^{2} \\ 
\left| c_{13}\right| ^{2} & \left| c_{12}\right| ^{2} & \left| c_{11}\right|
^{2} & \left| c_{23}\right| ^{2} & \left| c_{22}\right| ^{2} & \left|
c_{21}\right| ^{2} & \left| c_{33}\right| ^{2} & \left| c_{32}\right| ^{2} & 
\left| c_{31}\right| ^{2} \\ 
\left| c_{33}\right| ^{2} & \left| c_{32}\right| ^{2} & \left| c_{31}\right|
^{2} & \left| c_{23}\right| ^{2} & \left| c_{22}\right| ^{2} & \left|
c_{21}\right| ^{2} & \left| c_{13}\right| ^{2} & \left| c_{12}\right| ^{2} & 
\left| c_{11}\right| ^{2} \\ 
\left| c_{23}\right| ^{2} & \left| c_{22}\right| ^{2} & \left| c_{21}\right|
^{2} & \left| c_{13}\right| ^{2} & \left| c_{12}\right| ^{2} & \left|
c_{11}\right| ^{2} & \left| c_{33}\right| ^{2} & \left| c_{32}\right| ^{2} & 
\left| c_{31}\right| ^{2} \\ 
\left| c_{12}\right| ^{2} & \left| c_{11}\right| ^{2} & \left| c_{13}\right|
^{2} & \left| c_{22}\right| ^{2} & \left| c_{21}\right| ^{2} & \left|
c_{23}\right| ^{2} & \left| c_{32}\right| ^{2} & \left| c_{31}\right| ^{2} & 
\left| c_{33}\right| ^{2} \\ 
\left| c_{32}\right| ^{2} & \left| c_{31}\right| ^{2} & \left| c_{33}\right|
^{2} & \left| c_{22}\right| ^{2} & \left| c_{21}\right| ^{2} & \left|
c_{23}\right| ^{2} & \left| c_{12}\right| ^{2} & \left| c_{11}\right| ^{2} & 
\left| c_{13}\right| ^{2} \\ 
\left| c_{22}\right| ^{2} & \left| c_{21}\right| ^{2} & \left| c_{23}\right|
^{2} & \left| c_{12}\right| ^{2} & \left| c_{11}\right| ^{2} & \left|
c_{13}\right| ^{2} & \left| c_{32}\right| ^{2} & \left| c_{31}\right| ^{2} & 
\left| c_{33}\right| ^{2}
\end{array}
\right]  \notag \\
&&
\end{eqnarray}
This payoff is for the general matrix given in eq. (\ref{genmatrix}). In
case an exchange of strategies by Alice and Bob also exchanges their
respective payoffs the game is said to be symmetric. The idea of
evolutionary stability in mathematical biology is generally considered in
symmetric contests. In a symmetric contest payoff to a player is then
defined by his strategy and not by his identity. Payoffs in classical mixed
strategy game can be obtained from eq. (\ref{payoff}) when initial state is $%
\left| \psi _{in}\right\rangle =\left| 11\right\rangle $ and the game is
symmetric when $\alpha _{ij}=\beta _{ji}$ in the matrix (\ref{genmatrix}).
Similarly the quantum game played using the general quantum state of eq. (%
\ref{IniStat}) becomes symmetric when $\left| c_{ij}\right| ^{2}=\left|
c_{ji}\right| ^{2}$ for all constants $c_{ij}$ in the initial quantum state
of eq. (\ref{IniStat}). This condition should hold along with the
requirement $\alpha _{ij}=\beta _{ji}$ on the matrix (\ref{genmatrix}). The
payoff to Alice or Bob i.e. $P_{A}$, $P_{B}$ then does not need a subscript
and we can use only $P$.

We now come to the question of evolutionary stability in this quantized
version of the RSP game.

\section{Evolutionary stability in the quantized RSP game}

We define a strategy by a pair of numbers $(p,p_{1})$ when players are
playing quantized RSP game. It is understood that the identity operator is
,then, applied with probability $1-p-p_{1}$. Similar to requirements in eq. (%
\ref{ess1}) and eq. (\ref{ess2}) the conditions making a strategy $(p^{\star
},p_{1}^{\star })$ an ESS can now be written as \cite{smith, weibull}

\begin{eqnarray}
\text{1. \ \ \ }P\{(p^{\star },p_{1}^{\star }),(p^{\star },p_{1}^{\star })\}
&>&P\{(p,p_{1}),(p^{\star },p_{1}^{\star })\}  \notag \\
\text{2. if }P\{(p^{\star },p_{1}^{\star }),(p^{\star },p_{1}^{\star })\}
&=&P\{(p,p_{1}),(p^{\star },p_{1}^{\star })\}\text{ then}  \notag \\
P\{(p^{\star },p_{1}^{\star }),(p,p_{1})\} &>&P\{(p,p_{1}),(p,p_{1})\}
\label{ESSconds}
\end{eqnarray}
Suppose $(p^{\star },p_{1}^{\star })$ is a mixed NE then

\begin{equation}
\left\{ \frac{\partial P}{\partial p}\mid _{\substack{ p=q=p^{\star }  \\ %
p_{1}=q_{1}=p_{1}^{\star }}}(p^{\star }-p)+\frac{\partial P}{\partial p_{1}}%
\mid _{\substack{ p=q=p^{\star }  \\ p_{1}=q_{1}=p_{1}^{\star }}}%
(p_{1}^{\star }-p_{1})\right\} \geq 0
\end{equation}
Using substitutions

\begin{equation}
\begin{array}{cc}
\left| c_{11}\right| ^{2}-\left| c_{31}\right| ^{2}=\bigtriangleup _{1}, & 
\left| c_{21}\right| ^{2}-\left| c_{11}\right| ^{2}=\bigtriangleup _{1}^{%
{\acute{}}%
} \\ 
\left| c_{13}\right| ^{2}-\left| c_{33}\right| ^{2}=\bigtriangleup _{2}, & 
\left| c_{22}\right| ^{2}-\left| c_{12}\right| ^{2}=\bigtriangleup _{2}^{%
{\acute{}}%
} \\ 
\left| c_{12}\right| ^{2}-\left| c_{32}\right| ^{2}=\bigtriangleup _{3}, & 
\left| c_{23}\right| ^{2}-\left| c_{13}\right| ^{2}=\bigtriangleup _{3}^{%
{\acute{}}%
}
\end{array}
\end{equation}
we get

\begin{eqnarray}
\frac{\partial P}{\partial p} &\mid &_{\substack{ p=q=p^{\star }  \\ %
p_{1}=q_{1}=p_{1}^{\star }}}=p^{\star }(\bigtriangleup _{1}-\bigtriangleup
_{2})\left\{ (\alpha _{11}+\alpha _{33})-(\alpha _{13}+\alpha _{31})\right\}
+  \notag \\
&&p_{1}^{\star }(\bigtriangleup _{1}-\bigtriangleup _{3})\left\{ (\alpha
_{11}+\alpha _{32})-(\alpha _{12}+\alpha _{31})\right\} -  \notag \\
&&\bigtriangleup _{1}(\alpha _{11}-\alpha _{31})-\bigtriangleup _{2}(\alpha
_{13}-\alpha _{33})-\bigtriangleup _{3}(\alpha _{12}-\alpha _{32}) \\
\frac{\partial P}{\partial p_{1}} &\mid &_{\substack{ p=q=p^{\star }  \\ %
p_{1}=q_{1}=p_{1}^{\star }}}=p^{\star }(\bigtriangleup _{3}^{%
{\acute{}}%
}-\bigtriangleup _{1}^{%
{\acute{}}%
})\left\{ (\alpha _{11}+\alpha _{23})-(\alpha _{13}+\alpha _{21})\right\} + 
\notag \\
&&p_{1}^{\star }(\bigtriangleup _{2}^{%
{\acute{}}%
}-\bigtriangleup _{1}^{%
{\acute{}}%
})\left\{ (\alpha _{11}+\alpha _{22})-(\alpha _{12}+\alpha _{21})\right\} + 
\notag \\
&&\bigtriangleup _{1}^{%
{\acute{}}%
}(\alpha _{11}-\alpha _{21})+\bigtriangleup _{2}^{%
{\acute{}}%
}(\alpha _{12}-\alpha _{22})+\bigtriangleup _{3}^{%
{\acute{}}%
}(\alpha _{13}-\alpha _{23})
\end{eqnarray}
For the matrix (\ref{matrix}) above equations can be written as

\begin{eqnarray}
\frac{\partial P}{\partial p} &\mid &_{\substack{ p=q=p^{\star }  \\ %
p_{1}=q_{1}=p_{1}^{\star }}}=\bigtriangleup _{1}\left\{ -2\epsilon p^{\star
}-(3+\epsilon )p_{1}^{\star }+(1+\epsilon )\right\} +  \notag \\
&&\bigtriangleup _{2}\left\{ 2\epsilon p^{\star }+(1-\epsilon )\right\}
+\bigtriangleup _{3}\left\{ (3+\epsilon )p_{1}^{\star }-2\right\} \\
\frac{\partial P}{\partial p_{1}} &\mid &_{\substack{ p=q=p^{\star }  \\ %
p_{1}=q_{1}=p_{1}^{\star }}}=\bigtriangleup _{1}^{%
{\acute{}}%
}\left\{ -p^{\star }(3-\epsilon )+2\epsilon p_{1}^{\star }+(1-\epsilon
)\right\} -  \notag \\
&&\bigtriangleup _{2}^{%
{\acute{}}%
}\left\{ 2\epsilon p_{1}^{\star }-(1+\epsilon )\right\} +\bigtriangleup
_{3}^{%
{\acute{}}%
}\left\{ (3-\epsilon )p^{\star }-2\right\}
\end{eqnarray}
Also the payoff difference in the second condition of an ESS and given in
eq. (\ref{ESSconds}) reduces to

\begin{eqnarray}
&&P\{(p^{\star },p_{1}^{\star }),(p,p_{1})\}-P\{(p,p_{1}),(p,p_{1})\}  \notag
\\
&=&(p^{\star }-p)[-\bigtriangleup _{1}\{2\epsilon p+(3+\epsilon
)p_{1}-(1+\epsilon )\}+  \notag \\
&&\bigtriangleup _{2}\{2\epsilon p+(1-\epsilon )\}+\bigtriangleup
_{3}\{(3+\epsilon )p_{1}-2\}]+  \notag \\
&&(p_{1}^{\star }-p_{1})[-\bigtriangleup _{1}^{%
{\acute{}}%
}\{(3-\epsilon )p-2\epsilon p_{1}-(1-\epsilon )\}-  \notag \\
&&\bigtriangleup _{2}^{%
{\acute{}}%
}\{2\epsilon p_{1}-(1+\epsilon )\}+\bigtriangleup _{3}^{%
{\acute{}}%
}\{(3-\epsilon )p-2\}]
\end{eqnarray}
With the substitutions $p^{\star }-p=x$ and $p_{1}^{\star }-p_{1}=y$ above
payoff difference is

\begin{eqnarray}
&&P\{(p^{\star },p_{1}^{\star }),(p,p_{1})\}-P\{(p,p_{1}),(p,p_{1})\}  \notag
\\
&=&\bigtriangleup _{1}x\left\{ 2\epsilon x+(3+\epsilon )y\right\}
-\bigtriangleup _{2}(2\epsilon x^{2})-\bigtriangleup _{3}xy(3+\epsilon )- 
\notag \\
&&\bigtriangleup _{1}^{%
{\acute{}}%
}y\left\{ 2\epsilon y-(3-\epsilon )x\right\} +\bigtriangleup _{2}^{%
{\acute{}}%
}(2\epsilon y^{2})-\bigtriangleup _{3}^{%
{\acute{}}%
}xy(3-\epsilon )  \label{2ndESS}
\end{eqnarray}
provided

\begin{equation}
\frac{\partial P}{\partial p}\mid _{\substack{ p=q=p^{\star }  \\ %
p_{1}=q_{1}=p_{1}^{\star }}}=0\text{ \ \ \ \ \ \ \ \ \ \ \ \ \ }\frac{%
\partial P}{\partial p_{1}}\mid _{\substack{ p=q=p^{\star }  \\ %
p_{1}=q_{1}=p_{1}^{\star }}}=0  \label{conds}
\end{equation}
The conditions in eq. (\ref{conds}) together define the mixed NE $(p^{\star
},p_{1}^{\star })$. Consider now the modified RSP game in classical form
obtained by setting $\left| c_{11}\right| ^{2}=1$ and all the rest of the
constants to zero. The eqs. (\ref{conds}) now become

\begin{eqnarray}
-2\epsilon p^{\star }-(\epsilon +3)p_{1}^{\star }+(\epsilon +1) &=&0  \notag
\\
(-\epsilon +3)p^{\star }-2\epsilon p_{1}^{\star }+(\epsilon -1) &=&0
\end{eqnarray}
and $p^{\star }=p_{1}^{\star }=\frac{1}{3}$ is obtained as a mixed NE for
all the range $-1<\epsilon <0$. From eq. (\ref{2ndESS}) we get

\begin{eqnarray}
&&P\{(p^{\star },p_{1}^{\star }),(p,p_{1})\}-P\{(p,p_{1}),(p,p_{1})\}  \notag
\\
&=&2\epsilon (x^{2}+y^{2}+xy)=\epsilon \left\{
(x+y)^{2}+(x^{2}+y^{2})\right\} \leq 0  \label{differ}
\end{eqnarray}
In the classical form of the RSP game, therefore, the mixed NE $p^{\star
}=p_{1}^{\star }=\frac{1}{3}$ is a NE but not an ESS because the second
condition of ESS given in eq. (\ref{ESSconds}) does not hold.

Define now a new initial state as follows

\begin{equation}
\left| \psi _{in}\right\rangle =\frac{1}{2}\left\{ \left| 12\right\rangle
+\left| 21\right\rangle +\left| 13\right\rangle +\left| 31\right\rangle
\right\}  \label{inistat}
\end{equation}
and use it to play the game instead of the classical game obtained from $%
\left| \psi _{in}\right\rangle =\left| 11\right\rangle $. The strategy $%
p^{\star }=p_{1}^{\star }=\frac{1}{3}$ still forms a mixed NE because the
conditions given by eq. (\ref{conds}) hold true for it. However the payoff
difference of eq. (\ref{2ndESS}) is now given below when $-1<\epsilon <0$,
and $x,y\neq 0$

\begin{eqnarray}
&&P\{(p^{\star },p_{1}^{\star }),(p,p_{1})\}-P\{(p,p_{1}),(p,p_{1})\}  \notag
\\
&=&-\epsilon \left\{ (x+y)^{2}+(x^{2}+y^{2})\right\} >0
\end{eqnarray}
Therefore, the mixed Nash equilibrium $p^{\star }=p_{1}^{\star }=\frac{1}{3}$
not existing as an ESS in the classical form of the RSP game becomes an ESS
when the game is quantized and played using the initial entangled quantum
state given by the eq. (\ref{inistat}).

Note that from eq. (\ref{payoff}) the payoff sum to Alice and Bob $%
P_{A}+P_{B}$ can be obtained for both classical mixed strategy game (i.e. $%
\left| \psi _{in}\right\rangle =\left| 12\right\rangle $) and the quantum
game played using the quantum state of eq. (\ref{inistat}). For the matrix (%
\ref{matrix}) we write these sums as $(P_{A}+P_{B})_{cl}$ and $%
(P_{A}+P_{B})_{qu}$ for classical mixed strategy and quantum games
respectively and find

\begin{equation}
(P_{A}+P_{B})_{cl}=-2\epsilon \left\{
(1-p-p_{1})(1-q-q_{1})+p_{1}q_{1}+pq\right\}
\end{equation}
and

\begin{equation}
(P_{A}+P_{B})_{qu}=-\left\{ \frac{1}{2}(P_{A}+P_{B})_{cl}+\epsilon \right\}
\end{equation}
In case $\epsilon =0$ both the classical and quantum games are clearly zero
sum. For our slightly modified version of the RSP game we have $-1<\epsilon
<0$ and both versions of the game become non zero sum.

\section{Discussion}

Game theoretical modelling of interactions between living organisms in
natural word has been developed mostly during the last three decades. Use of
matrix games is quite common in areas such as theoretical and mathematical
biology. The RSP game that we investigate in present paper is also played in
nature like many other games. Lizards in the Coast Range of California play
this game using three alternative male strategies locked in an ecological
never ending process from which there seems little escape. On the other hand
the recently developed quantum game theory has been shown to find
applications in quantum information \cite{werner}. Though there is no
evidence yet the possibility of quantum games being played at molecular
level was hinted by Dawkins \cite{dawkins}. Trying to find the relevance of
ideas from population biology in quantum settings is something that we call
an inspiration from Dawkins' ideas.

The possibility of quantum mechanics playing a more direct role in life than
binding together atoms has attracted much attention \cite{farhi, frohlich}.
Quantum mechanics `fast tracking' a chemical soup to states that are
biological and complex is an idea about which physicists from many areas
have expressed opinions and the debate still continues. Supersymmetry in
particle physics giving a unified description of fermions and bosons have
also been suggested to provide an explanation of coding assignments in
genetic code \cite{bashford}. Patel's idea of quantum dynamics having a role
in the DNA replication is another interesting suggestion \cite{patel}.
Quantum game theory \cite{eisert, meyer} can also have possibly interesting
contributions to make towards attempts to understand quantum mechanical role
in life.

Mathematical biologists have successfully developed mathematical models of
evolution, especially, after attention was diverted to game theoretical
models of evolution \cite{weibull} and the idea of an ESS became central in
evolutionary game theory. The central idea of evolution i.e. survival of the
fittest is now be formulated as a mathematical algorithm usually known as
replicator dynamic. We suggest recent progress in quantum game theory allows
evolutionary ideas to enter and have a role in situations generally believed
to lie in the domain of quantum mechanics. This combination of evolutionary
ideas in quantum settings is interesting from several perspectives. Quantum
considerations in the evolution of genetic code and genetic algorithms in
which replicators receive their payoffs via quantum strategies are two
questions \cite{iqbal2} where evolutionary ideas can be incorporated in
quantum game like situations. Another possible relevance is the competing
chemical reactions in life molecules treated as players in a game. A winning
chemical reaction corresponding to life hints a role of quantum mechanics
because quantum strategies have been recently shown to be more effective
their classical counterparts \cite{marinatto, eisert}.

Population approach borrowed from evolutionary game theory with its central
idea of an ESS combined with recent developments in quantum game theory
provides a new approach to certain questions relating to role of quantum
mechanics in life. The analysis of the RSP game from evolutionary point of
view is an example where `stability' comes to a classical NE when players
revert to quantum strategies. The `stability' is with respect to an invasion
by mutants appearing in small numbers. This stability of NE coming out of
quantization can have a relevance in all the three situations indicated
above.

\section{Conclusion}

We explored evolutionary stability in a modified Rock-Scissors-Paper quantum
game. We showed that a mixed strategy NE not an ESS in classical version of
the game can be made an ESS when the two player play instead a quantum game
by using a selected form of initial entangled state on which they apply
unitary operators in their possession. Quantum mechanics, thus, gives
stability to a classical mixed NE against invasion by mutants. Stability
against mutants for a mixed classical NE can be made to disappear in certain
types of three player symmetric games when players decide to resort to
quantum strategies \cite{iqbal2}. Stability against mutants in pair-wise
contests coming as a result of quantum strategies have been shown a
possibility for only pure strategies in certain type of symmetric games \cite
{iqbal}. Our results imply the selected method of quantization \cite
{marinatto} can bring stability against mutants to a classical mixed NE in
pair-wise symmetric contests when the classically available number of pure
strategies to a player is increased to three from two. A different behavior
is also observed of mixed NE from pure NE in relation to quantization.

\end{document}